\documentclass[12pt,preprint]{aastex}
\usepackage{graphicx}

\shorttitle{Active Region Moss}
\shortauthors{Durgesh Tripathi et al.}

\begin{document}

\title{Evidence of Impulsive Heating in Active Region Core Loops}

\author{Durgesh Tripathi, Helen E. Mason}
\affil{Department of Applied Mathematics and Theoretical Physics, University of
Cambridge, Wilberforce Road, Cambridge CB3 0WA, UK}
\and

\author{James A. Klimchuk} 
\affil{NASA Goddard Space Flight Center, Greenbelt, MD20771, USA }
\email{d.tripathi@damtp.cam.ac.uk}

\begin{abstract} Using a full spectral scan of an active region from the
Extreme-Ultraviolet Imaging Spectrometer (EIS) we have obtained
Emission Measure EM$(T)$ distributions in two different moss regions
within the same active region. We have compared these with
theoretical transition region EMs derived for three limiting cases,
namely \textit{static equilibrium}, \textit{strong condensation} and
\textit{strong evaporation} from \cite{ebtel}. The EM distributions
in both the moss regions are strikingly similar and show a monotonically 
increasing trend from  $\log T[\mathrm{K}]=5.15\mbox{--}6.3$. Using 
photospheric abundances we obtain a consistent EM distribution for all ions. 
Comparing the observed and theoretical EM distributions, we find that the observed
EM distribution is best explained by the \textit{strong
condensation} case (EM$_{con}$), suggesting that a downward enthalpy
flux plays an important and possibly dominant role in powering the
transition region moss emission. The downflows could be due to
unresolved coronal plasma that is cooling and draining after having
been impulsively heated. This supports the idea that the hot loops
(with temperatures of 3{--}5~MK) seen in the core of active regions
are heated by nanoflares.
\end{abstract}


\section{Introduction}

Solar coronal heating has been one of the most intensely studied
problems in solar physics over the past 60 years. Despite major
advances in the observational and theoretical tools, the solution
remains elusive \citep[see e.g.,][]{klimchuk_2006}.  Active regions
provide a target of opportunity to study the dominant heating
mechanisms for the corona. The basic building blocks of active
regions are coronal loops. Because of the lack of cross-field
transport of mass and thermal energy in the corona, the problem of
coronal heating can effectively be reduced to a clear understanding
of the heating mechanism in a single isolated coronal loop or loop
strand.

The large-scale 1~MK loops and fan-like loops (which appear at the
periphery of active regions) are seen clearly in observations
recorded using the Transition Region and Coronal Explorer
\citep[TRACE;][]{trace} 171~{\AA} passband. Their properties appear
to be consistent with impulsive heating models \citep[see e.g.,][and references 
cited therein]{tripathi_2009, warren_2003, klimchuk_2006,
klimchuk_2009}. However, the 3{--}5~MK loops in the core of the
active regions are more diffuse and more difficult to isolate even
with present day observatories. Hence it is more difficult to
measure the physical plasma parameters along these hot core loops
and to compare those properties with simulated results
\citep[although see][]{lopez_2010}. The "moss" regions are located
at footpoints of hot (3{--}5~MK) loops \citep{martens, antiochos,
tripathi_submitted} and they map the photospheric magnetic flux very
closely \citep{tripathi_2008}. The moss regions, being the footpoints
of the hot loops, provide a unique opportunity to study the dominant
heating mechanism at work in active region core loops.

A number of publications involving observational results, analytical
and 1D numerical simulations have proposed that steady heating is at
play in the hot loops associated with moss regions.  The main
argument is that minimal variability is observed in brightness of
the moss emission or in the Doppler shifts and line widths of
spectral lines. \citet{antiochos} pointed out that impulsive heating
would produce transient loops in the cores of active regions that
would be seen by {\it TRACE} as the plasma cools through 1 MK.  Such
loops are present but not ubiquitous. The authors also noted that
the brightness of moss is quite steady when averaged over several
{\it TRACE} pixels.  \citet{brooks_warren} later showed that Doppler shifts and line widths
 observed by the Extreme ultraviolet Imaging Spectrometer (EIS) on
 {\it Hinode} are also quite steady. Finally, \cite{tripathi_submitted} measured
the density and thermal properties of moss observed by EIS and found
that they do not change significantly over periods ranging from an
hour to 5 days.

These observations provide compelling evidence that heating in the
cores of active regions must be quasi-steady {\it if}~the
cross-field scale of the heating is comparable to the 2 arcsec (1.5
Mm) resolution of EIS or larger. That is, we can conclude that the
heating is steady, but only if the plasma does not have unresolved
sub-structure. There is good reason to believe that such
sub-structure may exist, however \citep[e.g.,][]{klimchuk_2006}.
Individual thin strands could be highly time variable, and yet an
unresolved bundle would appear steady as long as the strands are not
in phase. This is true for both the coronal portions of the strands
and their moss footpoints.

We therefore seek a diagnostic of heating that does not require that the 
fundamental structures be resolved.  Doppler shifts are one possibility.  
Steady heating tends to produce no Doppler shift because the loop evolves 
to a static equilibrium \citep[although see][]{mariska_1983, klimchuk_2010}. 
Impulsive heating would produce a net red shift on the other hand.  In an 
unresolved bundle of strands, some strands
would have upflows due to evaporation and others would have
downflows due to condensation (cooling and draining).  Simulations
show that the upflows are faster, fainter, and shorter lived than
the downflows \citep{patsourakos_2006}.  Hence, the composite line
profile would be red shifted by several kilometers per second, and very
hot lines would also have a faint blue wing enhancement. The 
red shifts are predicted to decrease with temperature and may even become
blue shifts in the hottest lines \citep[][in preparation]{steve_jim}. The
observational picture concerning Doppler shifts in moss regions is
unclear. Some published measurements indicate minimal red-shifts
consistent with steady heating \citep[e.g.,][]{brooks_warren}, while
other measurements indicate larger red-shifts consistent with
impulsive heating \citep[e.g.,][]{doschek_1976, brueckner_1981, klimchuk_1987,
doschek_2008, delzanna_2008}. This is an important area of future
work.

Another diagnostic of coronal heating that works when there is
unresolved sub-structure is the emission measure distribution,
EM$(T)$, or the closely related differential emission measure
distribution, DEM$(T)$.  Since the plasma is roughly isothermal
along the coronal portion of a strand, EM$(T)$ in the corona is
determined primarily by the distribution of strand temperatures.  An
isothermal corona with all strands having the same temperature
implies steady heating \citep{warren_1999, landi_2002}, but some
coronal EM$(T)$ support a nanoflare interpretation 
\citep[e.g.,][]{patsourakos_2009, brendan, reale_2009}.
\cite{warren_2010} suggested that the lack of a strong correlation
between hot and warm emission in the core of an active region is
evidence that the heating must be steady. We disagree that 
this is the only interpretation. The relative intensities of coronal emission 
formed at, say, $\log T$ = 6.0 and 6.5 can vary greatly depending on the magnitude of the
nanoflares \citep[e.g.,][Figs. 4 and 7, dashed curves]{ebtel}.

Moss is the transition region footpoints of hot strands, and the
EM$(T)$ there is determined largely by the variation of temperature
along the strands.  \cite{ebtel} have shown that the form of EM$(T)$
is different under three limiting conditions: \textit{static
equilibrium}, \textit{strong condensation} and \textit{strong
evaporation}.  We make use of these differences in the work reported
here.  We measure the EM distribution of two moss regions observed
by EIS and compare them to the theoretical predictions.  In this way
we distinguish between steady and impulsive heating.

The rest of the paper is structured as follows. In section~\ref{obs}
we present our observations. In section~\ref{trems} we provide the
emission measure expressions following \cite{ebtel}.
Section~\ref{res} provides results and we summarise and discuss our
results in section~\ref{summ}.

\section{Observations} \label{obs}

\begin{figure}
\centering
\includegraphics[width=0.7\textwidth]{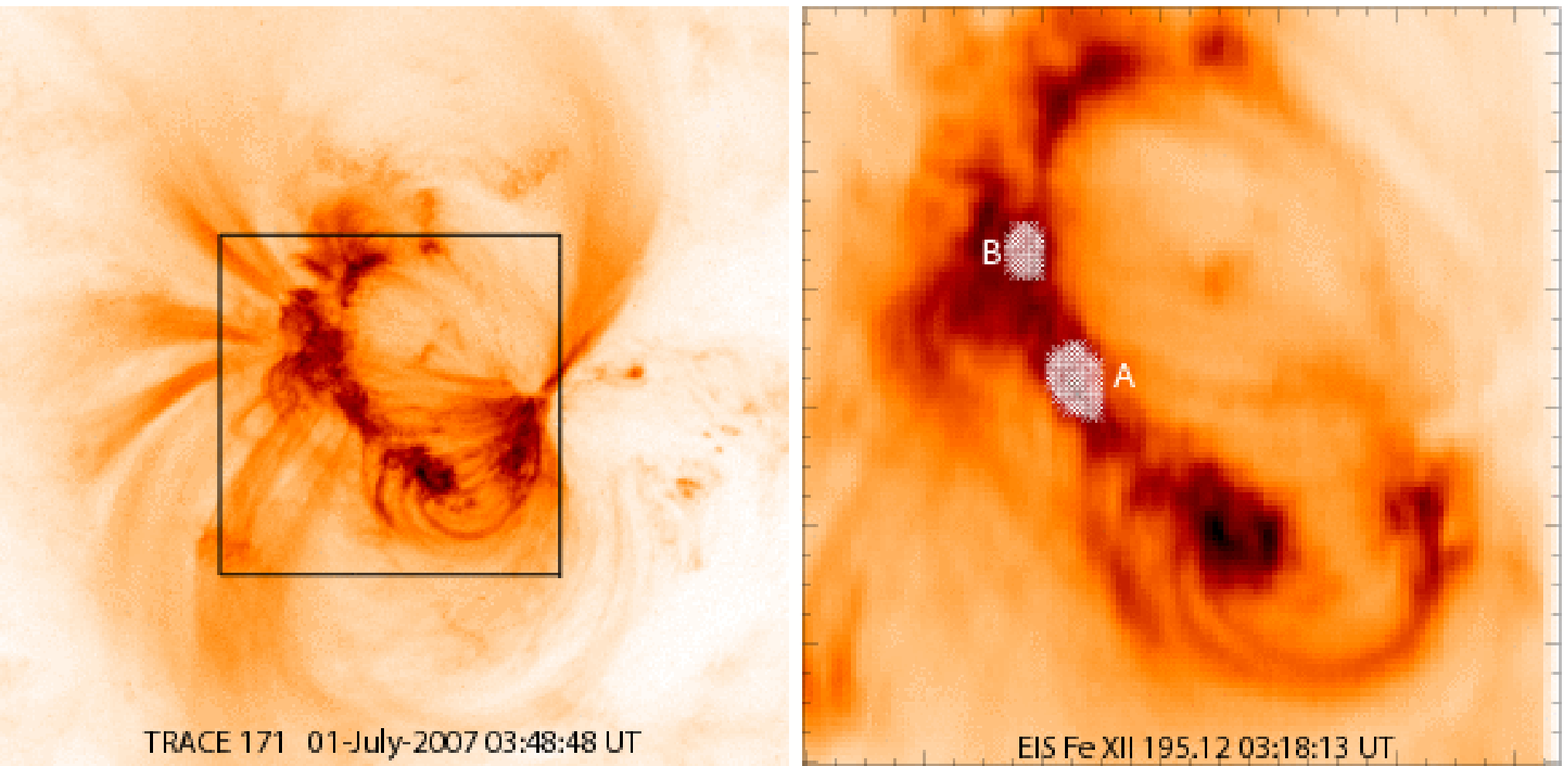}
\caption{Left panel: TRACE~171~{\AA} image showing the complete active
region. The over-plotted box shows the EIS field of view. Right panel: An
image obtained in \ion{Fe}{12}~195.12~{\AA}. The two regions labelled
'A' and 'B' were selected for detailed study. \label{context}}
\end{figure}

In this study, we have analysed observations recorded by the EIS aboard
Hinode. The EIS instrument observes the spectral ranges 170{--}210 and
250{--}290~{\AA}, providing observations in a broad range of temperatures
($\log T[\mathrm{K}]=4.7\mbox{--}7.3$). For detailed information on EIS see
\cite{eis}. For most of the EIS studies only a limited number of spectral lines
were chosen due to the limitations on telemetry. However, occasionally full
spectral scans of specific regions have been telemetered. Here we use one
such raster of an active region AR 10961 recorded on July 01, 2007 at
03:18 UT. The active region first appeared on the east limb on June 26th, 2007.
The data analysed were obtained from a single raster that scanned a field of view (FOV)
of 128{\arcsec} by 128{\arcsec} with the 1{\arcsec} slit, moving from west to east
with an exposure time of 25~sec. The over-plotted box on the {\it TRACE}~171~{\AA}
image displayed in the left panel of Fig.~\ref{context} marks the region which was
rastered by EIS. An EIS image obtained in \ion{Fe}{12}~$\lambda$195.12~{\AA} is
shown in the right panel of Fig.~\ref{context}.

The full spectral scan allowed us to choose spectral lines formed
over a broad range of temperature, which is essential in order to
obtain a good EM distribution of the plasma. This is the first time a full spectral 
scan is being used to derive an
EM distribution in active region moss. In our study we have used
relatively clean spectral lines (see Table~\ref{table}) with
temperatures of formation $\log T[\mathrm{K}]=5.1\mbox{--}6.5$ based on EIS spectral 
atlases \citep{ly_2009, yl_2009}. The
standard EIS software provided in \textsl{SolarSoft} were used to
process the data and \textsl{eis\_auto\_fit}, also provided in
\textsl{SolarSoft}, was used for Gaussian fitting the spectral
lines. To de-blend some of the blended lines labelled 'b'  in
Table~\ref{table} we have used the same procedure as in
\citet{tripathi_submitted} based on the suggestions of
\cite{peter_artb, peter_dens}.

\begin{table}
\centering
\caption{Spectral lines used to study the emission measure distribution in moss regions.\label{table}}
\begin{tabular}{lcc}
Ion     &Wavelength     & log~T\\
        & [{\AA}]           & [K] \\
\hline
\ion{O}{4}      & 279.6         & 5.15 \\
\ion{O}{4}      & 279.9     & 5.15 \\
\ion{O}{5}      & 248.5         & 5.35 \\
\ion{O}{6}      & 183.9         & 5.45 \\
\ion{O}{6}      & 184.1     & 5.45 \\
\ion{Mg}{5}     & 276.6     & 5.45 \\
\ion{Si}{6}     & 246.0     & 5.60 \\
\ion{Fe}{8}     & 185.2     & 5.60 \\
\ion{Mg}{6}     & 269.0         & 5.65 \\
\ion{Mg}{7}$^b$     & 278.4     & 5.80 \\
\ion{Si}{7}     & 275.3     & 5.80 \\
\ion{Fe}{9}     & 171.1     & 5.85 \\
\ion{Fe}{10}        & 184.5     & 6.05 \\
\ion{Si}{9}     & 258.1     & 6.05 \\
\ion{Si}{10}        & 261.0     & 6.15 \\
\ion{Fe}{11}        & 180.4     & 6.15 \\
\ion{Fe}{11}        & 188.3     & 6.15 \\
\ion{Fe}{12}        & 192.4     & 6.20 \\
\ion{Fe}{12}        & 193.5     & 6.20 \\
\ion{Fe}{12}$^b$    & 195.1     & 6.20 \\
\ion{Fe}{13}        & 202.0     & 6.25 \\
\ion{Fe}{14}$^b$    & 274.2     & 6.30 \\
\ion{Fe}{15}        & 284.2     & 6.35 \\
\ion{Fe}{16}        & 263.0     & 6.45 \\
\ion{Ca}{14}        & 193.8     & 6.55\\
\hline
\end{tabular}
\end{table}

We have derived EM following the approach of \cite{pottasch} whereby
individual emission lines yield estimates of the emission measure at the
lines' temperatures of formation. An EM distribution is built up by considering
lines formed over a wide range of temperatures. The method requires the
contribution function to be approximated by a function such that it is defined
to be a constant over the temperature range $\log\,T_{\rm max}-0.15$ to
$\log\,T_{\rm max}+0.15$ where $T_{\rm max}$ is the temperature where
the contribution function has its maximum. A detailed description of this
method is given in \cite{tripathi_submitted}.

\section{Theoretical Transition Region Emission Measure} \label{trems}

\cite{ebtel} derived the differential emission measure (DEM) distribution of the
transition region of a strand in the limiting cases of strong evaporation (Equation~A3),
strong condensation (Equation~A7), and static equilibrium (Equation~A12). The
transition region is taken here to be the region of steep gradient at the
strand footpoints.  Its temperature ranges from a chromospheric value up to
approximately one-half of the maximum temperature in the strand. Thus, it can reach
several MK for strands which are very hot. In the case of static equilibrium, a downward
heat flux from the corona powers the radiative losses from the transition region. These
losses are ignorable in the case of strong evaporation, where a downward heat flux
drives an evaporative upflow. The energy flux of the upflow is dominated by enthalpy
if the flow is subsonic. For the final case of strong condensation, the enthalpy of a
downflow powers the transition region radiation, with the heat flux being relatively
unimportant. See \cite{steve_2008} and  \cite{steve_peter, steve_peter2} for detailed
discussions of the importance of downflows in heating the transition region and cooling the
corona.

\begin{figure}
\centering
\includegraphics[width=0.7\textwidth]{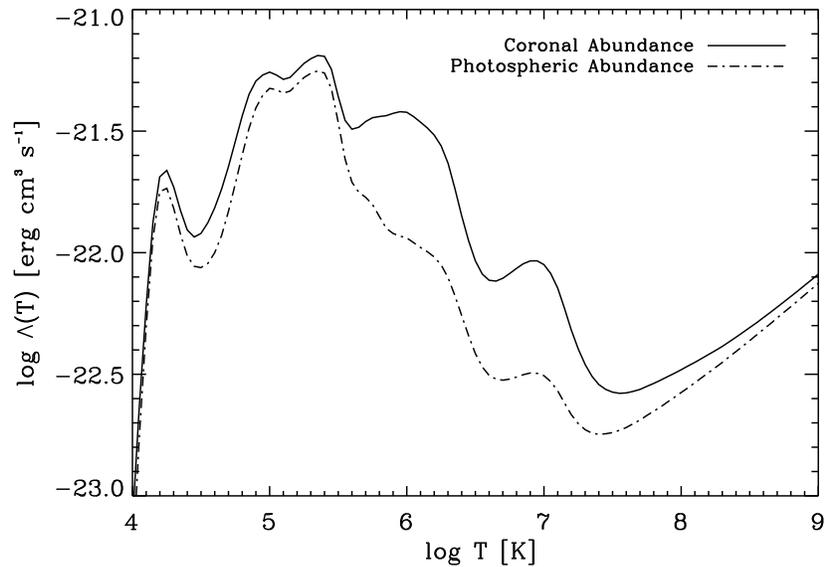}
\caption{Radiative energy losses using the photospheric abundances  (dashed-dotted line) of
\cite{photo_abund}, coronal abundances (solid line) of \cite{coronal_abund}  and CHIANTI~v6.0 ionization equilibrium
\cite{chianti_v6}. \label{rad_loss}}
\end{figure}

The expressions given in \cite{ebtel} are for the differential
emission measure.  The conversion to emission measure (EM) requires that
they be multiplied by temperature and by $ln(10)$, giving

\begin{equation} \label{stat}
\centering
EM_{se}\approx~ln(10)~\left(\frac{\kappa_0}{14}\right)^{1/2}\frac{\bar{P}~T^{3/4}}{k~\Lambda(T)^{1/2}}~...~static~equilibrium
\end{equation}

\begin{equation} \label{cond}
\centering
EM_{con} \approx{-}~ln(10)~\frac{5~k~J_0~T}{\Lambda(T)} ~...~strong~condensation
\end{equation}

\begin{equation} \label{eva}
\centering
EM_{ev} \approx \frac{ln(10)}{20}~\frac{\kappa_0}{k^3}~\frac{\bar{P}^2~T^{1/2}}{J_0}~...~strong~evaporation
\end{equation}

\noindent where $\kappa_0$=1.0$\times$10$^{-6}$ in cgs units,
\textit{k} is Boltzmann's constant, \textit{T} is temperature. \textit{J$_0$}
is mass flux defined as \textit{J$_0$ = nv}, where \textit{n} and
\textit{v} are electron number density and plasma flow speed
respectively. $\bar{P}$ is average pressure along a strand.
$\Lambda(T)$ is the optically thin radiative loss function.
Fig.~\ref{rad_loss} displays the radiative losses as a function of
temperature calculated with CHIANTI \citep{chianti_v1} by free-free,
radiative recombination and by line radiation for photospheric
abundances (dashed-dotted line) of \cite{photo_abund} and coronal
abundances (solid line) of \cite{coronal_abund}. For this
calculation, we have used the CHIANTI~v6.0.1 ionization equilibrium
\citep{chianti_v6}. The above equations give the emission measure
curves for individual loop strands. While computing the theoretical
EMs to compare with the observed EMs, we have considered the average
pressure ($\bar{P}$) and mass flux $J_0$ as arbitrary constants. We
have restricted ourselves to the temperature dependence of
EM$_{ev}$, EM$_{con}$ and EM$_{se}$ and
compared the shapes of the theoretical EM curves with those derived
observationally as explained in section~\ref{obs}. The assumption of
constant pressure and constant mass flux in a given strand is
justified so long as the different parts of the solar atmosphere are
connected along the magnetic field. However, this assumption would
break down if plasma in different layers of the atmosphere were
physically in different structures as was suggested by e.g.,
\cite{landi_feldman} and references therein.

\section{Results} \label{res}

\begin{figure}
\centering
\includegraphics[width=0.7\textwidth]{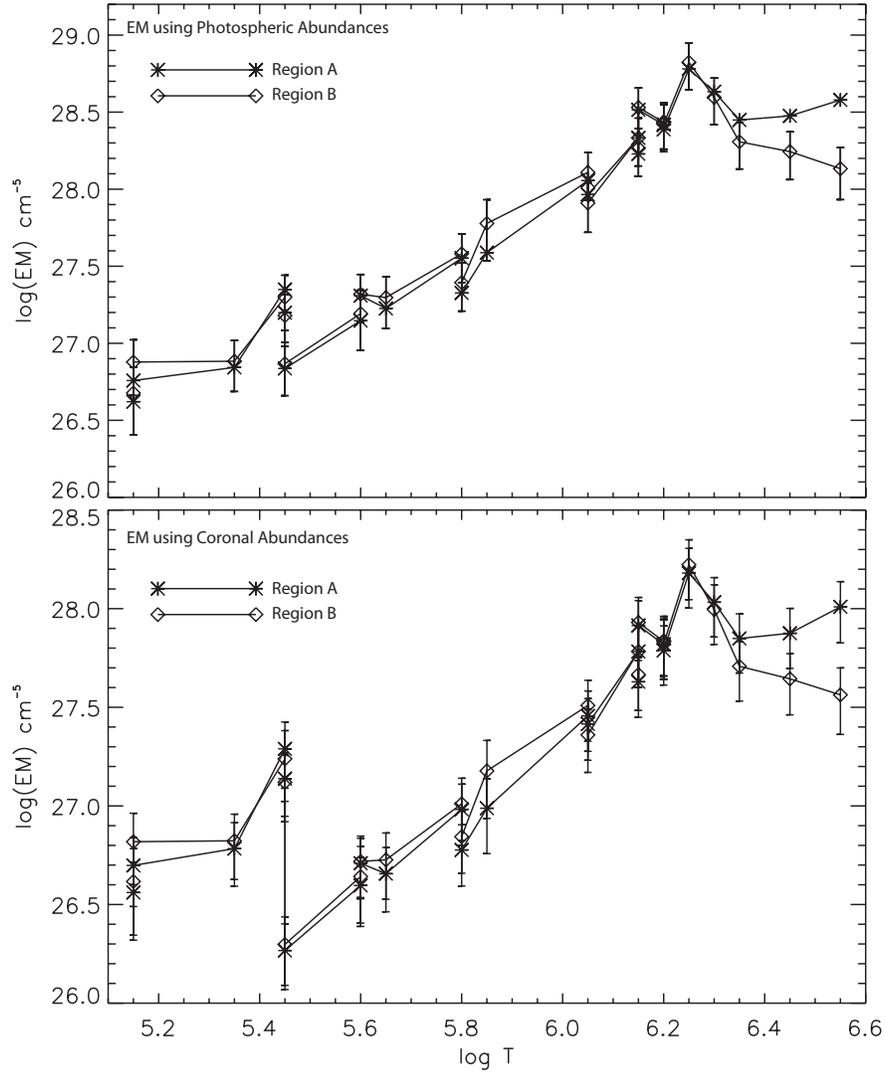}
\caption{Observed EM(T) for two different moss regions A and B shown in
the right panel of Fig.~\ref{context} using the photospheric abundances (top panel) of
\cite{photo_abund} and the coronal abundances (bottom panel) of
\cite{coronal_abund}.\label{obs_em}}
\end{figure}

Figure~\ref{obs_em} displays EM curves for two different moss regions
labelled "A" and "B" in the right panel of Fig.~\ref{context}. The EM curves
are created using the photospheric abundances (top panel) of \cite{photo_abund}
and the coronal abundances (bottom panel) of \cite{coronal_abund}.  The error
bars on the EMs are computed by considering a 30\% error in the observed
intensities for all the spectral lines, which is a rough estimate of the cumulative
errors due to radiometric calibration, line fitting and atomic calculations.

The EM curves for the two different regions displayed in
Fig.~\ref{obs_em} are very similar to each other when the same
abundances are used. However, for a given region the EM curves are
significantly different for the two different abundances, mainly at
lower temperatures. The difference in the trend of the EM curves is
basically due to the three data points on the extreme left,
corresponding to three Oxygen lines namely \ion{O}{4}, \ion{O}{5}
and \ion{O}{6}. The EMs for Oxygen lines do not appear to be
consistent with the EMs for other lines when coronal abundances are
used. However, this discrepancy disappears when we use photospheric
abundances. From the figure we note that the EMs of the Oxygen lines
do not change when coronal or photospheric abundances are used.
However, for other lines, the EMs increased by a factor of four for
photospheric abundances, which makes the EMs for other lines
consistent with the EMs for the Oxygen lines. Therefore, it
seems appropriate to conclude that the composition in these moss
regions is photospheric. However, we note that Oxygen is the 
only element in our dataset with a high first ionization potential (FIP). 
The high-FIP \ion{O}{6} lines agree with the low-FIP \ion{Mg}{5} line 
at $\log T = 5.4$ only with photospheric abundances. To verify the 
conclusion that moss regions must have photospheric abundances, it is 
important to examine additional pairs of high-FIP and low-FIP lines 
formed at the same temperature.

The observational picture of abundances in the solar upper atmosphere is
not well established \citep[see][for a recent review]{feldman_widing_2003}. Using
observations recorded by Skylab SO82A, it was found that in newly emerging
active regions the composition was nearly photospheric. However,
within a few hours after their emergence the composition starts to modify and it
approaches a typical coronal composition \citep[see][]{sheeley, widing, widing_feldman}.
Using data from the Coronal Diagnostic Spectrometer (CDS) aboard the Solar and
Heliospheric Observatory (SoHO), \cite{peter_abund} found photospheric composition
in an emerging flux region. The active region studied here had emerged about 6 days
prior to the date when the data analysed here were recorded. Considering the results
obtained using the Skylab observations, our analysis should suggest coronal abundances.
It is worth mentioning, however, that Skylab SO82A was a slitless spectrometer which
produced overlapping spectroheliorams providing partial information for
active regions. \cite{giulio_abund} suggested that the previous results based on Skylab 
observations overestimated the FIP bias by a large factor for active regions, if the plasma 
is isothermal rather than multi-thermal. Furthermore, since the moss regions are 
presumably confined to a relatively low altitude in the solar atmosphere, the FIP effect might 
not be significant in these regions. Therefore, photospheric abundances could prevail in moss regions.

The EM curves for the moss regions show that the peak of the emission
comes from \ion{Fe}{13} ($\log T = 6.25$) as suggested by
\cite{tripathi_submitted}. In addition, the EM$(T)$ obtained 
using photospheric abundances  shows a monotonic increase at lower temperatures 
from $\log T = 5.15$. However, the EM$(T)$ obtained using coronal abundances 
shows a minimum in the emission measure at $\sim \log T = 5.5$ similar to the 
EM curve obtained using the DEM distribution derived from SERTS EUV spectrum using the same abundances 
\citep{brosius_1996, brosius_2000}.

\begin{figure}
\centering
\includegraphics[width=0.85\textwidth]{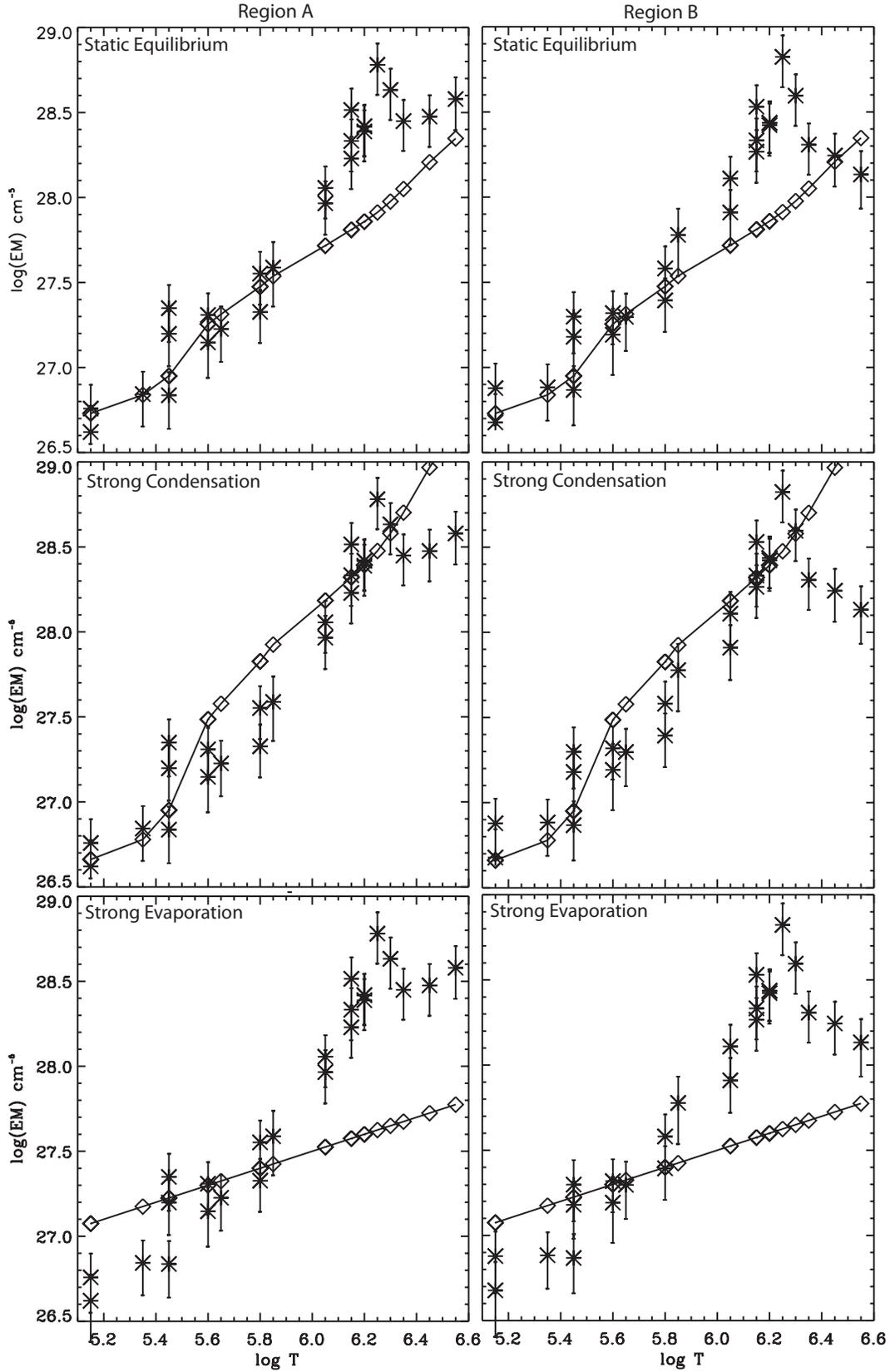}
\caption{Theoretical (solid lines with over-plotted diamonds) EM curve
for three limiting cases using Eqns .1, 2 \& 3 and observed (asterisks)
EM curves using photospheric abundances of \cite{photo_abund}.\label{em_photo_abund}}
\end{figure}

We have compared the observed EM distributions obtained using
coronal and photospheric abundances with those predicted
theoretically for the three limiting cases of \textit{static equilibrium
(EM$_{se}$, Eq.~\ref{stat})}, \textit{strong condensation} (EM$_{con}$,
  Eq.~\ref{cond}) and \textit{strong evaporation} (EM$_{ev}$,
  Eq.~\ref{eva}) for both the regions 'A' and 'B' marked in the right
panel of Fig.~\ref{context}.  Fig.~\ref{em_photo_abund} displays the
observational EMs obtained using photospheric abundances
over-plotted with the theoretical EMs for the three limiting cases,
both for region 'A' and region 'B' in the left and right columns respectively.

The theoretical curve that provides the best overall
match with the observed points is EM$_{con}$ i.e., the strong
condensation case (see plots in the middle row in
Fig.~\ref{em_photo_abund}). The EM$_{se}$ for static equilibrium
matches quite well with observed EM for temperatures below 1~MK but
not at higher temperatures. However, that for strong condensation
(EM$_{con}$) matches quite well for $\log T =$ 5.2 -- 6.25. Above
$\log T = 6.3$, the EM$_{con}$ suggests an increasing trend, whereas
the observational EM starts to decrease. We can understand this
apparent discrepancy as follows.

Suppose that unresolved strands are heating impulsively to very high
temperatures by nanoflares and subsequently cooling.  As stated
earlier, the transition region in a given strand extends to roughly
one-half of the maximum strand temperature. Recently heated strands
will therefore contribute to moss emission over a wide range of
temperatures, while strands that have cooled will only contribute
emission at lower temperatures.  We can imagine a collection of
curves like those in the middle panels of Fig.~\ref{em_photo_abund}
except truncated at different temperatures above $\log T = 6.3$. The
sum of those curves would have a shape similar to the observed
EM$(T)$, rolling over at $\log T = 6.3$. This basic effect can be seen in the
transition region of model 4 in \citet[][Fig. 7, dot-dashed
curve]{ebtel}. The figure shows DEM$(T)$ (= EM$(T)/T$) and is
therefore flat rather than positively sloped at the lower
temperatures.  The temperature where EM$(T)$ is expected to roll
over will depend upon the number distribution and brightness of
strands in different stages of cooling. Strands which have cooled
below 1 MK should be relatively faint due to mass draining and
therefore will have a reduced contribution to the composite EM$(T)$.

The EM$_{ev}$ curve for strong evaporation does not show
any resemblance to the observed EM curve.  This is not surprising,
since, in the nanoflare scenario, evaporative upflows are much
fainter and much shorter lived than the subsequent condensation
downflows and therefore contribute minimally to the net emission.

When using coronal abundances, we did not find a reasonable
correspondence between the observed EM$(T)$ and any of the
theoretical EM$(T)$. However, this conclusion is based on the 
disagreement in the emission measure at $\log T < 5.45$, and therefore 
depends entirely on EM of Oxygen lines. The EM$_{con}$ and EM$_{se}$ 
curves fit the data equally well in this questionable range $\log T < 5.45$. 
However, we note that at $\log T > 5.45$ EM$_{con}$ provides a better fit 
to the data than EM$_{se}$ regardless of which abundance set is assumed.

\section{Summary and discussion} \label{summ}

We have obtained EM distributions in active region moss using full
spectral scans recorded with EIS for two different regions. The EM
distributions for the two regions show very good agreement with each
other. The EM obtained using coronal abundances does not give
consistent values for different lines. However, photospheric
abundances seem to provide consistent EMs. The EM distributions show
a monotonically increasing trend from $\log T =$~5.15~{--}~6.3. Of
the three theoretical limiting cases we have considered, strong
condensation (EM$_{con}$) best reproduces the observations. Static
equilibrium (EM$_{con}$) does a good job at lower temperatures, but
it fails significantly above $\log T = 6.0$.  Strong evaporation
(EM$_{con}$) is clearly inconsistent with the observations.  We
conclude from this that downflows play an important role in powering
the moss radiation.  A similar conclusion was reached by
\cite{steve_2008} on the basis of observed red-shifts \citep[see
e.g.,][]{doschek_1976, brueckner_1981, klimchuk_1987, doschek_2008, delzanna_2008,
  warren_2008, tripathi_2009}.

As we have discussed, one obvious source of downflows is the cooling
and draining of plasma that has been impulsively heated by
nanoflares. As long as the heated strands have a sub-resolution
diameter, this picture is consistent with observations of apparently
steady intensities, Doppler shifts, and line widths.  Another
possible source of downflows is material from Type II spicules. This
newly discovered phenomena appears to represent jets of
chromospheric plasma that is heated to coronal temperatures as it is
ejected \citep{depontieu}.  Presumably the material subsequently
cools and falls.  Much more observational and modeling work is
necessary before any firm conclusions can be drawn.

Although our results support the picture of nanoflares and possibly Type II 
spicules in the cores of active regions, we cannot rule out steady heating entirely.  
If magnetic flux strands constrict appreciably at their base, then EM$_{se}(T)$ will 
have a different shape from that which we have shown \citep[e.g.][]{warren_2010}. 
The theoretical curve could in principle be brought in line with observations.  
However, because the transition region is so thin, the constriction must be very 
dramatic and the transition region must fall right where the constriction occurs.  
This would seem rather surprising given that the height of the transition region 
depends on the coronal pressure and moves up and down in response to 
changes in the heating rate, either slow or rapid. It has been found that 
the area of quiet Sun network boundaries increases by a factor of about 1.6 
between $\log T = 5.4$ and $6.0$ \citep{patsourakos_1999}. How this relates to moss 
regions discussed here is unclear.

The result that photospheric abundances apply in regions of downflow
may hold important clues about the origin of the FIP effect, the
fractionation of elements based on their first ionization potential.
If fractionation (coronal abundances) occurs in regions of steady
heating, where conditions are static, but not in regions of
impulsive heating, where there are continual upflows and downflows,
then it is suggested that the fractionation process involves some
sort of long-term settling.  Further observational and modeling work
is needed also in this regard.

\acknowledgments{We thank an anonymous referee for carefully reading the 
manuscript and comments. DT and HEM acknowledge from STFC. The work of JAK
was supported by the NASA Living With a Star Program. We acknowledge the loops IV workshop as an opportunity to stimulate discussions and collaborate on this project.
We thank the CHIANTI consortium. We thank Dr Giulio Del Zanna for various discussions 
and Dr Peter Young for providing his fitting IDL routines
  in \textsl{Solarsoft}. Hinode is a Japanese mission developed and
  launched by ISAS/JAXA, collaborating with NAOJ as a domestic
  partner, NASA and STFC (UK) as international partners. Scientific
  operation of the Hinode mission is conducted by the Hinode science
  team organized at ISAS/JAXA. This team mainly consists of scientists
  from institutes in the partner countries. Support for the
  post-launch operation is provided by JAXA and NAOJ (Japan), STFC
  (U.K.), NASA, ESA, and NSC (Norway). }




\end{document}